\documentclass[12pt,a4paper]{article}
\usepackage{a4wide}
\usepackage{latexsym}
\usepackage{amssymb}
\usepackage{amsmath, cite}
\usepackage{color}

 \makeatletter
\@addtoreset{equation}{section} \makeatother

\def\bfone{\relax{\rm 1\kern-.35em 1}}

 \makeatletter
\@addtoreset{equation}{section} \makeatother

\def\be {\begin{equation}}
\def\ee {\end{equation}}
\def\bea {\begin{eqnarray}}
\def\eea {\end{eqnarray}}
\def\bc {\begin{center}}
\def\ec {\end{center}}
\def\a  {\alpha}

\def\D  {\Delta}

\def\bfg {\begin{figure}}
\def\efg {\end{figure}}
\def\bi {\begin{itemize}}
\def\ei {\end{itemize}}
\def\nn {\nonumber}

\def\la {\label}
\def\le {\left}
\def\ri {\right}

\DeclareFontFamily{U}{rsf}{} \DeclareFontShape{U}{rsf}{m}{n}{
  <5> <6> rsfs5 <7> <8> <9> rsfs7 <10-> rsfs10}{}
\DeclareMathAlphabet\Scr{U}{rsf}{m}{n}

\begin{document}

\begin{center}
{\bf \large{ Modified Newton's Law of Gravitation Due to Minimal Length in Quantum Gravity \\[10mm]}}
\large Ahmed Farag
Ali$^{\star}$\footnote{Electronic address~:~\upshape{ahmed.ali@fsc.bu.edu.eg~;~ahmed.ali@uleth.ca }} and
A. Tawfik$^{
\dagger \ddagger}$\footnote{\upshape{Electronic address~:~a.tawfik@eng.mti.edu.eg~;~ atawfik@cern.ch}}\\[7mm]
\small{$^\star$Dept. of Physics, Faculty of Sciences, Benha University, Benha, 13518, Egypt.}\\[4mm]
\small{$^{\dagger}$Egyptian Center for Theoretical Physics (ECTP), MTI University, Cairo, Egypt.}\\[4mm]
\small{$^\ddagger$Research Center for Einstein Physics, Freie-University Berlin, Berlin, Germany.}
\end{center}

\begin{abstract}

A recent theory about the origin of the gravity suggests that the
gravity is originally an entropic force. In this work, we discuss the
effects of generalized uncertainty principle (GUP) which is
proposed by some approaches to \emph{quantum gravity} such as string
theory, black hole physics and doubly special relativity theories (DSR),
on the area law of the entropy. This leads to a $\sqrt{\text{Area}}$-type
correction to the area law of entropy which imply that the number
of bits $N$ is modified. Therefore, we obtain a modified
Newton's law of gravitation. Surprisingly, this modification
agrees with different sign with the prediction of Randall-Sundrum II model
which contains one uncompactified extra dimension. Furthermore,
such modification may have observable consequences
at length scales much larger than the Planck scale.

\end{abstract}




\section{Introduction}
\par\noindent

The earliest idea about the connection between gravitation
and the existence of a fundamental length was
proposed in \cite{mead}. In the last two decades,
the existence of a minimal length is one of the most
interesting predictions of some approaches related to quantum
gravity such as String Theory, Black hole physics and
non-commutative geometry \cite{guppapers,BHGUP,AmelinoCamelia:2008wq}.
The existence of a minimal length is considered as a consequence
of the string theory because strings obviously can not interact
at distances smaller than the string size. Furthermore, the black hole
physics suggests that the uncertainty relation should be modified
near the Planck energy scale due to the fact that the photons emitted
from the black hole suffers from two major errors; the first one is the
error by Heisenberg classical analysis and the second
one is because the black hole mass varies during the emission
process and the radius of the horizon changes accordingly
\cite{guppapers,BHGUP}. An interesting measure
gedanken experiment was proposed in \cite{Scardigli} involving
micro-black holes at the Planck scale of spacetime which leads to the GUP.
This independent model depends on Heisenberg principle and Schwarzschild radius.
Recently, It was found that polymer quantization suggests
the existence of minimal length in similar way to string theory
and black hole physics \cite{polymer}. Therefore, all these
different approaches suggests that the standard uncertainty
relation in quantum mechanics is modified to yield
{\it Generalized Uncertainty Principle} (GUP)\cite{guppapers,BHGUP,polymer}.
In light of this, such modifications can play an essential role
as the quantum gravitational corrections which would open an interesting window
for quantum gravity phenomenology
\cite{qgc,gupp3a,Scardigli:1995qd,gupp3b,gupp3c,
 gupp3d,dvprl,dvcjp,sabine,kmm,kempf,brau}.


In a one-dimensional chain as the Ising model, when assuming
that every single spin is positioned at a distance $d$ apart
from the two neighborhoods. Then, the macroscopic state of such
a chain is defined by $d$. The entire chain would have various
configurations so that  if $d\rightarrow l$, the chain has much
less configurations than if $d \ll l$, where $l$ is the chain's
length. Statistically, the entropy is given by the number of
microscopic states $S=k_B\ln \Omega$. Due to second law of
thermodynamics, such a system tends to approach a state of
maximal entropy so that the chain in the macroscopic state $d$
tends to go to a $d$ state with a much higher entropy. The force
that causes such a statistical tendency is defined as the entropic
force. In light of this, the entropic force is a phenomenological
mechanism deriving a system to approach maximum entropy i.e.,
increasing the number of microscopic states that will be inhered
in the system's phase space. There are various examples on the
entropic force, for example polymer molecules and the elasticity of rubber bands.

Recently, Verlinde proposed  that the gravity is not fundamental
force and can be considered as an entropic force \cite{everlind1}.
The earliest idea about gravity that is regarded as a non-fundamental
interaction has been introduced by Sakharov \cite{sakharoov67},
where the spacetime background is assumed to emerge as a mean
field approximation of underlying microscopic degrees of freedom.
Similar behavior is observed  in hydrodynamics \cite{hydro1}.
It is found that the entropy of black hole is related to the
horizon's area at the black hole's horizon, while the temperature
is related to the surface gravity. Both entropy  and temperature
are assumed to be related to the mass of black hole \cite{bhtherm}.
Thus, the connection between thermodynamics and geometry leads
to Einstein's equations of gravitational field from relations
connecting heat, entropy, and temperature \cite{jack}.
The Einstein's equations  connect energy-momentum tensor with space
geometry. Advocating the gravity as non-fundamental interaction
leads to the assumption that gravity would be explained as an
entropic force caused by changes in the information associated
with the positions of material bodies \cite{everlind1}.
When combining the entropic force with the Unruh temperature,
the second law of Newton is obtained. But when combining it
with the holographic principle and using the equipartition
law of energy, the Newton's law of gravitation is obtained.
It was investigated in \cite{Zhang:2010hi} modification  of the entropic force
due to corrections to the area law of entropy which is derived from quantum effects and
extra dimensions.

Apart from the controversial debate on the origin of gravity
\cite{Kobakhidze:2010mn,Hossenfelder:2010ih}, we investigate the
impact of GUP on the entropic force and derive essential quantities including
potential modification to the Newton's law of gravity.

There were some studies for the effect of some versions of GUP
on the Newton's law of gravity in \cite{Nozari:2011gj}. Also, Non-Commutative Geometry which
is considered as a completely planck scale effect  has  been studied to derive the modified
Newton's law of gravity \cite{Nicolini:2010nb,Bastos:2010au,Nozari:2011et,Mehdipour:2012nj}.
All these approaches for studying the
Planck scale effects on the Newton's law of gravity are based on
the following scheme: ~ modified theory of gravity $\rightarrow$
modified black hole entropy$\rightarrow$
modified holographic surface entropy $\rightarrow$
Newton's law corrections. We followed the same scheme in our paper
using the new version of GUP proposed in \cite{advplb, Ali:2010yn,Das:2010zf}, and
we a got a new corrections in our current work, which are distinct from the previous studies.
Moreover, we compared our results with Randall-Sundrum
model of extra dimension which also predicts
the modification of Newton's law of gravity at the planck scale \cite{Randall:1999vf, potential},
where we think there may be some connection between generalized uncertainty principle and extra dimension
theories because they predicting similar physics at least for the case of Newton's law of
gravity which may be considered as a distinct result from the previous studies.

The present paper is organized as follows. Section \ref{sec:gup}
reviews briefly the generalized uncertainty principle
that was proposed in \cite{advplb, Ali:2010yn,Das:2010zf} . Section
\ref{sec:entropicf} is devoted to review the entropic force
and  gravitational interaction \cite{everlind1}. The effect of utilizing
GUP impact on the entropic force is introduced in
section \ref{sec:efgup}. In section \ref{sec:correct},
we estimate the GUP correction to the Newton's law of gravitation.
The conclusions are given in section \ref{sec:conc}.

\section{The Generalized Uncertainty Principle}
\label{sec:gup}

It is conjectured that the standard commutation relations
at short distances would be modified.
A new form of GUP was proposed \cite{advplb,Ali:2010yn,Das:2010zf}
and found consistent with the Doubly Special Relativity (DSR) theories,
the string theory and the black holes physics.
It predicts a maximum observable momentum and a minimal measurable length.
With satisfying Jacobi identity, GUP is found to ensure the
relations $[x_i,x_j]=0=[p_i,p_j]$:

\bea
[x_i, p_j]\hspace{-1ex} &=&\hspace{-1ex} i \hbar\hspace{-0.5ex}
\left[  \delta_{ij}\hspace{-0.5ex} - \hspace{-0.5ex} \alpha \hspace{-0.5ex}
\left( p \delta_{ij} + \frac{p_i p_j}{p} \right) + \alpha^2 \hspace{-0.5ex}
\left( p^2 \delta_{ij}  + 3 p_{i} p_{j} \right) \hspace{-0.5ex} \right], \label{eq:alfaa}
\label{comm01}
\eea
where $\alpha = {\alpha_0}/{M_{p}c} = {\alpha_0 \ell_{p}}/{\hbar}$
and $M_{p} c^2$ stand for Planck energy. $M_{p}$ and $\ell_{p}$
is Planck mass and length, respectively. $\alpha_{0}$ sets
on the upper and lower bounds to $\alpha$.


For a particle having an energy scale comparable
to the Planck's one, the physical momentum would be a subject of
a  modification \cite{advplb,Ali:2010yn,Das:2010zf}

\bea
p_i = p_{0i} \left(1 - \alpha p_0 + 2\alpha^2 p_0^2 \right), \label{mom1}
\eea

where $x_i = x_{0i}$ and $p_{0j}$ satisfy the canonical commutation
relations $ [x_{0i}, p_{0j}] = i \hbar~\delta_{ij}$ . Here, $p_{0i}$
can be interpreted as the momentum at low energies and $p_{i}$
as that at high energies, and the variable $p_0$ is
the value of the momentum at the low energy scales.

This newly proposed GUP suggests that the space is quantized
into fundamental units which may be the Planck length.
The quantization of space has been shown within the
context of loop quantum gravity in \cite{LQG}. \\

In a series of earlier papers, the effects of GUP  was
investigated on atomic, condensed matter, preheating phase of the universe
systems, black holes at LHC\cite{Ali:2011fa, Chemissany:2011nq,Ali:2012mt}, the weak equivalence
principle (WEP), the Liouville theorem (LT) in statistical
mechanics \cite{faragali}. It was found that the GUP can potentially
explain the small observed violations of the WEP in neutron
interferometry experiments\cite{exp} and also predicts a
modified invariant phase space which is relevant to the
Liouville theorem. It was derived in \cite{Ali:2011fa} the first bound
for $\alpha_0$ is about  $\sim10^{ 17}$,
which would approximately gives $\alpha\sim 10^{-2}~$GeV$^{-1}$.
The other bound  of $\alpha_0$  which is $\sim10^{10}$.
This bound means that $\alpha\sim10^{-9}~$GeV$^{-1}$.
As discussed in \cite{Tawfik:2012hz}, the exact bound on
$\alpha$ can be obtained by comparison with observations and
experiments. It seems that the gamma rays
burst would allow us to set an upper value
for the GUP-charactering parameter $\alpha$ which we would like
to report on this in the future.\\

Recently, it has been suggested in \cite{Nature} that
the GUP implications can be measured directly in quantum
optics laboratories which definitely confirm the theoretical predictions
given in \cite{Ali:2011fa,dvprl}. Definitely, this is considered
as a milestone in the road of quantum gravity phenomenology.\\

In section \ref{sec:entropicf}, we  briefly review the
assumptions that the gravitational force
would be originated from an entropic nature.

\section{Gravity as an entropic force}
\label{sec:entropicf}

Recently, Erick Verlinde \cite{everlind1} has utilized
Sakharov's proposal \cite{sakharoov67} that the gravity
would not be considered a fundamental force. Concretely,
it was suggested that the gravitational force might be
originated to an entropic nature. As discussed in the
introduction, this assumption is based on the relation
between the gravitation and thermodynamics \cite{bhtherm}.
According to thermodynamics and holographic principle,
Verlinde's approach results in the Newton's law. Moreover,
the Friedmann equations can also be derived \cite{Cai:2010hk}.
At temperature $T$, the entropic force $F$ of  a gravitational
system is given as
\be
F \Delta x = T \Delta S, \la{force}
\ee
where $\Delta S$ is the change in the entropy so
that at a displacement $\Delta x$,
each particle carries its own portion of entropy change.
From  the correspondence between the entropy change
$\Delta S$ and the information
about the boundary of the system and using Bekenstein''s
argument \cite{bhtherm}, it is assumed that $ \Delta S = 2 \pi k_B$,
where $\Delta x = \hbar/m$ and $k_B$ is the Boltzmann constant.
\be
\Delta S = 2 \pi k_B \frac{m c}{\hbar} \Delta x, \label{entropy}
\ee
where $m$ is the mass of the elementary component $c$ is speed
of light and $\hbar$ is the Planck constant, respectively.

Turning to the holographic principle which assumes that
for any closed surface without worrying
about its geometry inside, all physics can be represented
by the degrees of freedom on this surface itself. This implies
that the quantum gravity can be described by a topological
quantum field theory, for which all physical degrees of freedom
can be projected onto the boundary\cite{thooft}. The information
about the holographic system is given by  $N$ bits forming an ideal gas.
It is conjectured that $N$ is proportional to the entropy of the holographic screen,

\bea
N= \frac{4 S}{k_B},\la{Nbits}
\eea
then according to Bekenstein's entropy-area relation \cite{bhtherm}
\be
S=\frac{k_B c^3}{4 G \hbar} A.
\ee
Therefore, one gets
\be
N=\frac{A c^3}{G \hbar} = \frac{4 \pi r^2 c^3}{G \hbar},
\ee
where $r$ is the radius of the gravitational system and the
area of the holographic screen $A= 4 \pi r^2$ is implemented
in deriving this equation.
It is assumed that each bit emerges outwards from the
holographic screen i.e., one dimension. Therefore each bit
carries an energy equal to $k_B T/2$, so using the equipartition
rule to calculate the energy of the system, one gets
\be
E=\frac{1}{2} N k_B T = \frac{2 \pi c^3 r^2}{G \hbar} k_B T = M c^2. \la{energy}
\ee
By substituting Eq. (\ref{force}) and Eq. (\ref{entropy}) into Eq. (\ref{energy}), we get
\be
F=G \frac{M m}{r^2},
\ee
making it clear that Newton's law of gravitation can be derived
from first principles.

In section \ref{sec:efgup}, we study the effect of GUP-approach
on the entropic force and hence on the Newton's law of gravitation.

\section{GUP impact on the black hole thermodynamics}
\label{sec:efgup}

Taking into consideration the GUP-approach
\cite{advplb,Ali:2010yn,Das:2010zf},
and because black holes are considered as a good laboratories
for the clear connection between thermodynamics and gravity,
the black hole thermodynamics will be analyzed in this section.
Furthermore, how the entropy would be affected shall be
investigated, as well. In Hawking radiation, the emitted
particles are mostly photons and standard model (SM) particles.
From kinetic theory of gases, let us assume that gatherings
or clouds of points in the velocity space are  equally
spread in all directions. There is no reason that particles
would prefer to be moving in a certain direction. Then,
the three-moments are simply equal

\be p_1\approx p_2\approx p_3, \la{sym}
\ee
leading to
\bea
p^2&=& \sum_{i=1}^{3} p_i p_i \approx 3~ p_i^2\,, \nn\\
\langle p_i^2 \rangle &\approx& \frac{1}{3}~ \langle p^2
\rangle\,.\la{3n} \eea
%
%
%
%
In order to find a relation between $\langle p^2 \rangle$ and $ \D p^2$,
we assume that the black hole behaves like a black body, while it
emits photons. Therefore, from Wien's law, the temperature
corresponding to the peak emission is given by

\bea
c\, \langle p \rangle =  2.82\, T_H.
\eea
We should keep in mind that the numerical factor, $2.82$, should be modified
by the grey-body factors which arise due to the spacetime
curvature around the black hole, but for simplicity we are just ignoring
this modification.

From Hawking uncertainty proposed by Scardigli in \cite{Scardigli:1995qd}
and Adler {\it et al.}  \cite{Adler}, the Hawking's temperature reads
\be T_H= \frac{1}{\pi}\, c\, \D p = \frac{1}{2.82}\, c\, \langle p \rangle. \ee
With  the relation $\langle p^2
\rangle = \D p^2 + {\langle p \rangle}^2$, we get
\bea
\langle p \rangle &=& 2.82\, \frac{1}{\pi}\, \D p = \sqrt{\mu}\, \D p \nn,\\
\langle p^2 \rangle &=& (1+\mu)\, \D p^2, \la{arg}
\eea
where $\mu= (2.82/\pi)^2$. Again, the parameter $\mu$ is modified if we consider
the grey-body factors which arise due to the spacetime
curvature around the black hole.

In order to have a corresponding inequality for Eq.\ (\ref{comm01}), we can utilize the arguments given in
 \cite{Cavaglia:2003qk}. Then
\be
\D x \D p \geq
 \frac{\hbar}{2}\le[1- \a \langle p \rangle- \a \langle
 \frac{p_i^2}{p} \rangle
+\a^2 \langle p^2 \rangle + 3 \a^2 \langle p_i^2\rangle \ri]\,.
\la{ineq}
\ee
It is apparent that implementing the arguments given in Eqs.\ (\ref{3n}) and (\ref{arg}) in the inequality gievn in Eq. (\ref{ineq}) leads to
\be \D x \D p \geq \frac{\hbar}{2} \le[1- \a_0 ~\ell_p~
\le(\frac{4}{3}\ri)~\sqrt{\mu}~~ \frac{\D p}{\hbar}+ ~2~
(1+\mu)~ \a_0^2 ~\ell_p^2 ~ \frac{\D p^2}{\hbar^2} \ri]\,,
\la{ineqII} \ee
 The resulting inequality, Eq. (\ref{ineqII}), is the only one that follows from Eq.\ (\ref{comm01}). Solving it  as a quadratic equation in $\D p$ results in
\be \frac{\D p}{\hbar}\geq\frac{2 \D
x+\a_0
~\ell_p~\le(\frac{4}{3}~\sqrt{\mu}~\ri)}{4~(1+\mu)~\a_0^2~\ell_{p}^2}\le(1-
\sqrt{1-\frac{8~(1+\mu)~\a_0^2\ell_{p}^2} {\le(2 \D x+\a_0
\ell_p\le(\frac{4}{3}\ri) ~\sqrt{\mu}~\ri)^2}}\ri)
\,.\la{gupso} \ee
The negatively-signed solution is considered as the one that refers to the standard uncertainty relation as $\ell_p/\D x \rightarrow 0$. Using the Taylor expansion, we obviously find that
\be
\Delta p \geq \frac{1}{\Delta x} \le(1- \frac{2}{3}\a_0 \ell_p \sqrt{\mu} \frac{1}{\Delta x} \ri).
\ee

Because the energy change reads $\D E\approx c~\D p$ and according to Scardigli in\cite{Scardigli:1995qd} and Adler
{\it et al.} \cite{Adler}, one can define the uncertainty in the energy $\D E$ as the energy carried away from the black hole through the emitted photon. In the following, we implement the procedure introduced in \cite{Eliasentropy}. We utilize the GUP-approach  \cite{advplb,Ali:2010yn,Das:2010zf} together with the assumptions given in Eqs.\ (\ref{3n}) and (\ref{arg}). Then, using natural units that  $\hbar=c=1$
\be
\D E \geq \frac{1}{\Delta x} \le(1- \frac{2}{3}\a_0 \ell_p \sqrt{\mu} \frac{1}{\Delta x} \ri). \la{energy1}
\ee

For a black hole absorbing a quantum particle with energy $E$ and size R, the area of the black hole is supposed to increase by the amount \cite{Areachange}.
\be
\Delta A \geq 8 \pi\, \ell_p^2\, E\, R,
\ee
The quantum particle itself implies the existence of finite bound given by
\be
\Delta A_{min} \geq 8 \pi\, \ell_p^2\, E\, \Delta\, x. \la{Darea}
\ee
Using Eq. (\ref{energy1}) in the inequality (\ref{Darea}), we obtain,
\be
\Delta A_{min} \geq 8 \pi \ell_p^2\le[ 1- \frac{2}{3}\a_0 \ell_p \sqrt{\mu} \frac{1}{\Delta x}\ri]. \la{Area}
\ee
According to the argument given in \cite{Eliasentropy}, the length scale is chosen to be
the inverse surface gravity
\be
\Delta x= 2\, r_s,
\ee
where $r_s$ is the is the Schwarzschild radius. This argument implies that
\be
(\Delta x)^2 \sim \frac{A}{\pi}. \label{DX}
\ee
Substituting  Eq. (\ref{DX}) into Eq. (\ref{Area}), we got
\be
\Delta A_{min}= \lambda  \ell_{p}^2 \le[1- \frac{2}{3}\, \a_0\, \ell_p\, \sqrt{\frac{ \mu \, \pi}{A}}\ri],
\ee
where parameter $\lambda$  will be fixed later. According to \cite{bhtherm}, the black hole's entropy is conjectured to depend on the horizon's area. From the information theory \cite{Adami:2004mx}, it has been found that the minimal increase of entropy should be independent on the area. It is just
one "bit" of information which is $b = \ln(2)$.
\be
\frac{dS}{dA}= \frac{\Delta S_{min}}{\Delta A_{min}} = \frac{b}{\lambda  \ell_{p}^2 \le[1- \frac{2}{3}\, \a_0\, \ell_p\,  \sqrt{\mu\, \pi \, A}\ri]},
\ee
where $b$ is a parameter. By expanding the last expression in orders of $\a$ and then integrating it, we get the entropy
\be
S= \frac{b}{\lambda \ell_p^2}\le[ A+ \frac{4}{3}\, \a_0\, \ell_p\, \sqrt{\mu\, \pi\, A}\ri].
\ee
Using Hawking-Bekenstein assumption, which relates entropy with the area, the value of constants $b/ \lambda= 1/4$, so that
\be
S=\frac{A}{4\, \ell_p^2} + \frac{2}{3}\, \a_0\,  \sqrt{\pi\, \mu\, \frac{A}{4\, \ell_p^2}}. \la{correctENTROPY}
\ee

Although it was found in \cite{Hod:2004cd} that the power--law corrections to
Bekenstein--Hawking area–entropy are ruled out based on arguments from Boltzmann--Einstein formula,  but
the it was found that the power-law corrections may explain the observed cosmic acceleration today \cite{Wang:2005bi}.

We conclude that the entropy is directly related to the area and gets a correction when applying GUP-approach. The temperature of the black hole is
\bea
T&=& \frac{\kappa}{8 \pi} \frac{dA}{dS}
=\frac{\kappa}{8 \pi} \le[1- \frac{2}{3}\, \a_0\, \ell_p\, \sqrt{\mu\, \frac{\pi}{A}} \ri].
\eea
So far, we conclude that the temperature is not only proportional to the surface gravity but
also it depends on the black hole's area.

\section{Modified Newton's law of gravitation due to GUP}
\label{sec:correct}

In this section we study the implications of the corrections
calculated for the entropy in  Eq. (\ref{correctENTROPY}),
and calculate how the number of bits of Eq. (\ref{Nbits})
would be modified which assume a new corrections
to the Newton's law of gravitation. Using the corrected entropy
given in Eq. (\ref{correctENTROPY}), we find that the number of bits
should also be corrected as follows.

\be
N^{\prime}= \frac{4 S}{k_B}= \frac{A}{\ell_{p}^2}+ \frac{4}{3}\, \a_0\, \sqrt{\mu\, \pi\, \frac{A}{\ell_{p}^2}}. \la{bits}
\ee
By substituting Eq. (\ref{bits}) into Eq. (\ref{energy}) and using Eq. (\ref{force}), we get
\be\label{eq:Ee}
E = 
      F\, c^2\, \le(\frac{r^2}{m\, G}+ \frac{\a\, \sqrt{\mu}\, r}{3\, m\, G}\ri).
\ee
It is apparent, that Eq. (\ref{eq:Ee}) implies a modification in the Newton's law of gravitation
\be
F= G \frac{M m}{r^2} \le(1- \frac{\a\, \sqrt{\mu}}{3\, r}\ri). \la{result}
\ee
This equation states that the modification in Newton's law of gravity seems to agree with the predictions of Randall-Sundrum II model \cite{Randall:1999vf} which contains one uncompactified extra dimension and length scale $\Lambda_R$. The only difference is the sign. The modification in Newton's  gravitational potential on brane \cite{potential} is given as
\bea
V_{RS} =\begin{cases} -G\frac{m M}{r} \le(1+\frac{4 \Lambda_R}{3 \pi r}\ri), &  r\ll\Lambda_R\\
& \\
-G\frac{m M}{r} \le(1+\frac{2 \Lambda_R}{3 r^2}\ri), & r\gg\Lambda_R
\end{cases}, \la{VRS}
\eea
where $r$ and $\Lambda_R$ are radius and the characteristic
length scale, respectively. It is clear that the gravitational
potential  is modified at short distance.
We notice that our result, Eq. (\ref{result}), agrees with different sign
with Eq. (\ref{VRS}) when  $r\ll\Lambda_R$. This result would say that
$\alpha \sim \Lambda_R$  which would help to set a new upper bound on the value of the parameter $\alpha$.
This means that the proposed GUP-approach \cite{advplb,Das:2010zf} is
apparently able to predict the same physics as Randall-Sundrum
II model. The latter assumes the existence of one extra dimension
compactified on a circle whose upper and lower halves are identified.
If the extra dimensions are accessible only to gravity and not to the
standard model field, the bound on their size can be fixed by an
experimental test of the Newton's law of gravitation, which has only been led
down to $\sim 4$ millimeter.  This was the result, about ten
years ago \cite{gt1}. In recent gravitational experiments,
it is found that the Newtonian gravitational force, the $1/r^2$-law,
seems to be maintained up to $\sim0.13-0.16~$mm \cite{gt2}. However,
it is unknown whether this law is violated or not at sub-$\mu$m range.
Further implications of this modifications have been discussed in
\cite{Buisseret:2007qd} which could be the same for the GUP
modification which is calculated in this paper. This similarly between the GUP implications and
extra dimensions implications would assume a new bounds on the GUP parameter $\alpha$ with respect to
the extra dimension length scale $\Lambda_R$.

\section{Conclusions}
\label{sec:conc}

In this paper, we tackle the consequences of the quantum gravity
on the entropic force approach which assumes a new origin of the gravitational force.
We found that the quantum gravity corrections lead to a modification
in the area law of the entropy which leads to a modification in the number of bits
$N$. According to Verlinde's theory of entropic force, the Newton's law of gravitation
would acquire new quantum gravity corrections due to the modified number of bits.
The modification in the Newton's Law of gravitation surprisingly agrees with the corrections predicted
by Randall-Sundrum II model with different sign. This would open a new
naturally arising question in our proposed research if the GUP
and  extra dimensions theories would predict the same physics.
We hope to report on this in the future.

\section*{Acknowledgments}
The research of AFA is supported by Benha University. The research of AT has been partly supported by the German--Egyptian Scientific Projects (GESP ID: 1378). AFA and AT like to thank Prof. Antonino Zichichi for his kind invitation to attend the International School of Subnuclear Physics 2012 at the ''Ettore Majorana Foundation and Centre for Scientific Culture'' in Erice-Italy, where the present work was started. The authors gratefully thank the anonymous referee
for useful comments and suggestions which helped to improve the paper.

%
%

\end{document}